\documentclass{sigchi-ext}
\usepackage[T1]{fontenc}
\usepackage{textcomp}
\usepackage[scaled=.92]{helvet} 
\usepackage{graphicx} 
\usepackage{balance}  
\usepackage{booktabs} 
\usepackage{ccicons}  
\usepackage{ragged2e} 

\def\plaintitle{One Button to Rule Them All} 

\def\emptyauthor{}
\def\plainkeywords{Button; haptic; modeling; simulation; tactility; force feedback; vibration; input device; haptic rendering; FD model; FDVV model.}

\title{Press'Em: Simulating Varying Button Tactility via FDVV Models}

\numberofauthors{6}
\author{%
  \alignauthor{%
    \textbf{Yi-Chi Liao}\\
    \affaddr{Aalto University} \\
    \affaddr{Helsinki, Finland}\\
    \email{yi-chi.liao@aalto.fi} }\alignauthor{%
    \textbf{Byungjoo Lee}\\
    \affaddr{KAIST}\\
    \affaddr{Daejeon, Republic of Korea}\\
    \email{byungjoo.lee@kaist.ac.krm} } \vfil \alignauthor{%
    \textbf{Sunjun Kim}\\
    \affaddr{Aalto University}\\
    \affaddr{Helsinki, Finland}\\
    \affaddr{KAIST}\\
    \affaddr{Daejeon, Republic of Korea}\\
    \affaddr{DGIST}\\
    \affaddr{Daegu, Republic of Korea}\\
    \email{sunjun.kim@aalto.fi} }\alignauthor{%
    \textbf{Antti Oulasvirta}\\
    \affaddr{Aalto University}\\
    \affaddr{Helsinki, Finland}\\
    \email{antti.oulasvirta@aalto.fi} }  }

\definecolor{linkColor}{RGB}{6,125,233}
\hypersetup{%
  pdftitle={\plaintitle},
  pdfauthor={\emptyauthor},
  pdfkeywords={\plainkeywords},
  bookmarksnumbered,
  pdfstartview={FitH},
  colorlinks,
  citecolor=black,
  filecolor=black,
  linkcolor=black,
  urlcolor=linkColor,
  breaklinks=true,
}

\copyrightinfo{\scriptsize Permission to make digital or hard copies of part or all of this work for personal or classroom use is granted without fee provided that copies are not made or distributed for profit or commercial advantage and that copies bear this notice and the full citation on the first page. Copyrights for third-party components of this work must be honored. For all other uses, contact the owner/author(s). \\
{\emph{CHI '20 Extended Abstracts, April 25--30, 2020, Honolulu, HI, USA.}} \\
\copyright~2020 Copyright is held by the author/owner(s). \\
ACM ISBN 978-1-4503-6819-3/20/04. \\
\url{http://dx.doi.org/10.1145/3334480.3383161}}

\begin{document}



\maketitle


\begin{abstract}
Push-buttons provide rich haptic feedback during a press via mechanical structures.
While different buttons have varying haptic qualities, few works have attempted to dynamically render such tactility, which limits designers from freely exploring buttons' haptic design. 
We extend the typical force-displacement (FD) model with vibration (V) and velocity-dependence characteristics (V) to form a novel \emph{FDVV} model.
We then introduce \emph{Press'Em}, a 3D-printed prototype capable of simulating button tactility based on FDVV models. 
To drive Press'Em, an end-to-end simulation pipeline is presented that covers (1) capturing any physical buttons, (2) controlling the actuation signals, and (3) simulating the tactility.
Our system can go beyond replicating existing buttons to enable designers to emulate and test non-existent ones with desired haptic properties. 
Press'Em aims to be a tool for future research to better understand and iterate over button designs.
  
\end{abstract}

\noindent\fbox{%
    \parbox{\columnwidth}{%
        \textbf{This demo accompanies a CHI '20 paper titled \\\emph{``Button Simulation and Design via FDVV Models''} \cite{button_design}.}
    }%
}

\keywords{\plainkeywords}

\ccsdesc[500]{Human-centered computing~Haptic devices}
\ccsdesc[500]{Human-centered computing~Interaction devices}

\section{Introduction}
\begin{marginfigure}[-7pc]
  \begin{minipage}{\marginparwidth}
    \centering
    \includegraphics[width=0.9\marginparwidth]{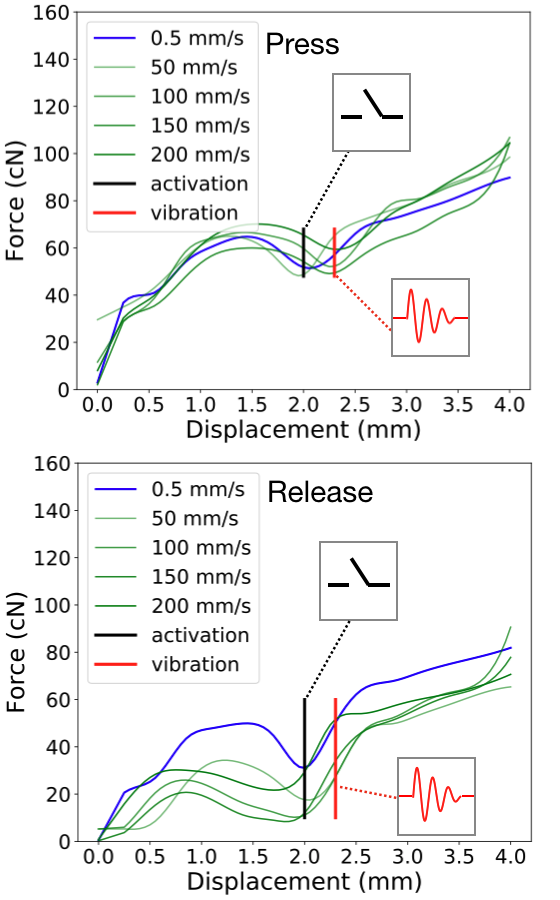}
    \caption{A force--displacement--vibration--velocity (FDVV) model represents speed-dependent physical responses of a button when pressed. 
  We show methods for capturing button presses as FDVV models, rendering them in a physical simulator, and editing and optimizing these in software. 
  The press and release models shown are for a 4 mm tactile button. Blue curves represent the corresponding (velocity-agnostic) force--displacement model typically measured by a probing machine with static and slow velocity.}~\label{fig:fdvvmodel}
  \end{minipage}
\end{marginfigure}

Each push-button provides unique haptic characteristics (tactility) during presses. 
Different haptic properties can lead to distinct experiences and users' performances; so much that gamers, programmers, and typists are willing to spend hundreds of dollars on keyboards just for the perfect tactility. 
Nonetheless, enhancing the tactile properties of buttons is tedious since testing different haptic profiles typically require complete design and engineering of physical buttons.
A notable exception is button-simulator~\cite{onebutton}, which models buttons using simple Force-Displacement (FD) Curves (see blue curves in Figure~\ref{fig:fdvvmodel}) and recreate force accordingly during a press. 
This approach falls short for three reasons: First, since a button is a spring-mass-damping system, the force rendered is a function depends on not just displacement but also velocity~\cite{velocity-fdcurve}. 
A single-FD model fails to capture the overall reality. 
Second, the structural vibration caused by fast tapping can not be recorded in FD models, either.
Lastly, during rendering, it applies no control for keeping the output forces to meet the references.

\begin{figure}[h]
  \centering
  \includegraphics[width=0.99\columnwidth]{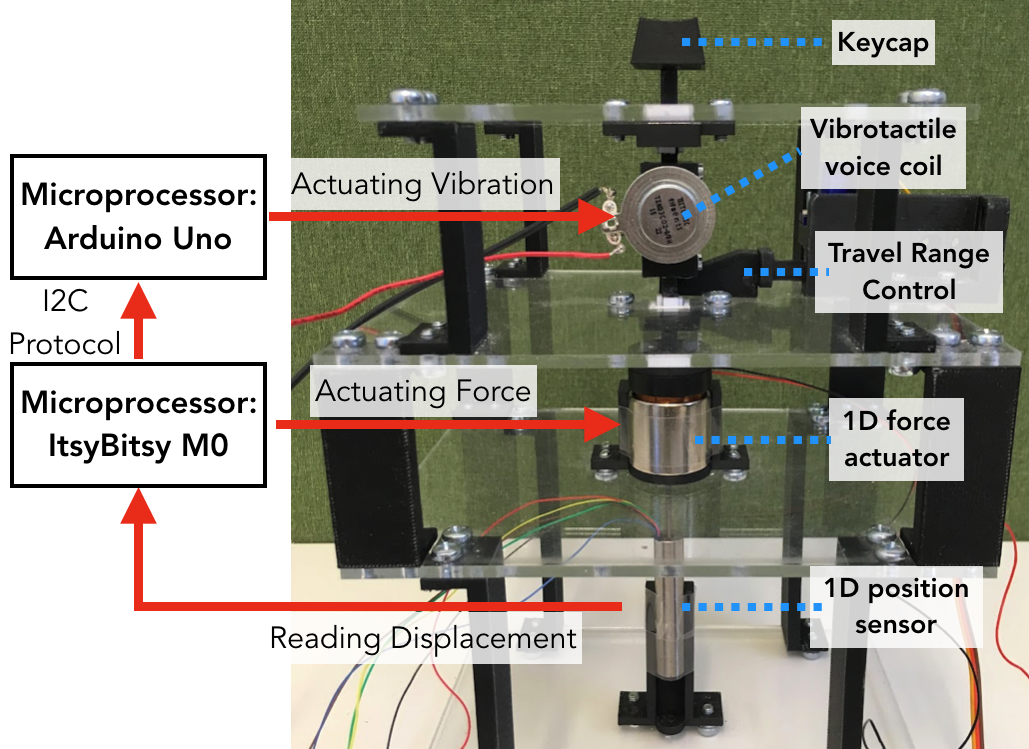}
  \caption{The simulator, Press'Em, includes a 1D sensor that tracks displacement, a 1D force actuator delivering various levels of forces, and a servo motor drives the travel-range control. These components are controlled by a microprocessor (ItsyBitsy M0). 
  The other microprocessor (Arduino Uno) drives a vibrotactile motor which is mounted near the keycap. }
  \label{fig:prototype}
\end{figure}

\begin{figure*}[t!]
  \centering
  \includegraphics[width=1.8\columnwidth]{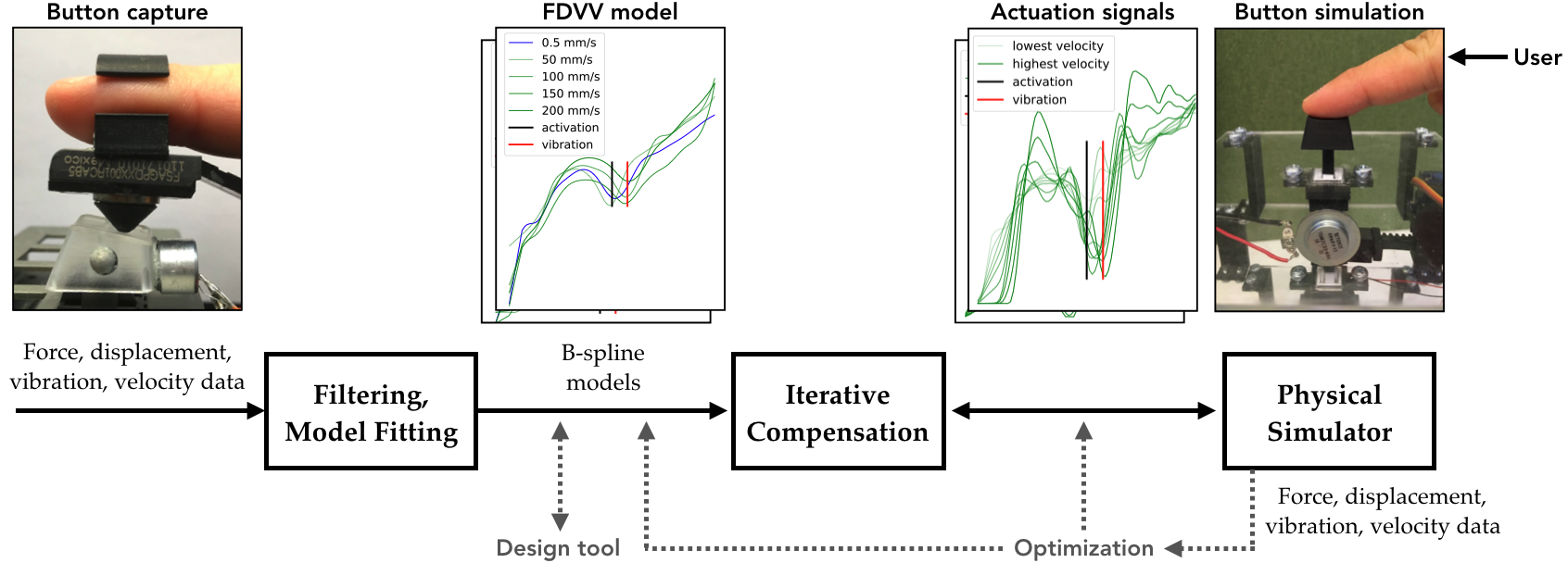}
  \caption{An end-to-end approach to button simulation. To capture an FDVV model of a button, sensors are placed on the finger, and the button is pressed multiple times. 
  The resulting force, displacement, vibration, and velocity data are filtered and modeled. 
  A designer can edit the model produced. 
  To render the model with a given physical plant, an iterative compensation process computes how to cancel the plant's own transfer function. 
  The resulting actuation signals drive the simulator.}~\label{fig:pipeline}
\end{figure*}

We introduce force--displacement--vibration--velocity (FDVV) model (Figure~\ref{fig:fdvvmodel}), which adds vibration response and velocity-dependence on top of the FD model. 
We further implemented Press'Em (Figure~\ref{fig:prototype}), a 3D-printed prototype that overcomes the aforementioned limitations by an end-to-end pipeline (Figure~\ref{fig:pipeline}) covering capture physical 
buttons, controlling, and simulation.
As a result, Press'Em provides designers with a complete platform for designing and testing different haptic profiles.

\section{Press'Em: the Button Simulator}

Press'Em is a physical simulator (in Figure \ref{fig:prototype}) capable of high-fidelity rendering of FDVV models.
Our first design goal was to provide high-frequency response and high-resolution rendering of forces and vibrations typical of buttons.
The second was to enable full control from the software side. 

\textbf{Sensors and actuators:} 
Figure~\ref{fig:prototype} presents the four main components: 
(1) a linear force actuator (Moticont HVCM-025-022-003-01), 
(2) a linear position sensor (LVDT MHR 250), 
(3) a voice coil acting as a vibrotactile motor (Tectonic Teax13C02-8), and 
(4) a servo motor (Tower Pro Micro Servo). 
The force actuator, the sensor, and the servo motor are controlled by an Adafruit ItsyBitsy M0 Express board. 
The vibrotactile voice coil is driven by an Arduino Uno board and wave shield (Adafruit Wave Shield for Arduino Kit). 
These two boards are connected via the I2C protocol. 
When adjustments to the \textit{overall travel range} are required, 
ItsyBitsy sends a command to the servo motor to adjust the location of the \textit{Travel Range Control}, which further alters the lowest reachable displacement of the \textit{Travel Range Limiter} and produces varying travel.

\begin{marginfigure}[-33pc]
  \begin{minipage}{\marginparwidth}
    \centering
    \includegraphics[width=0.99\marginparwidth]{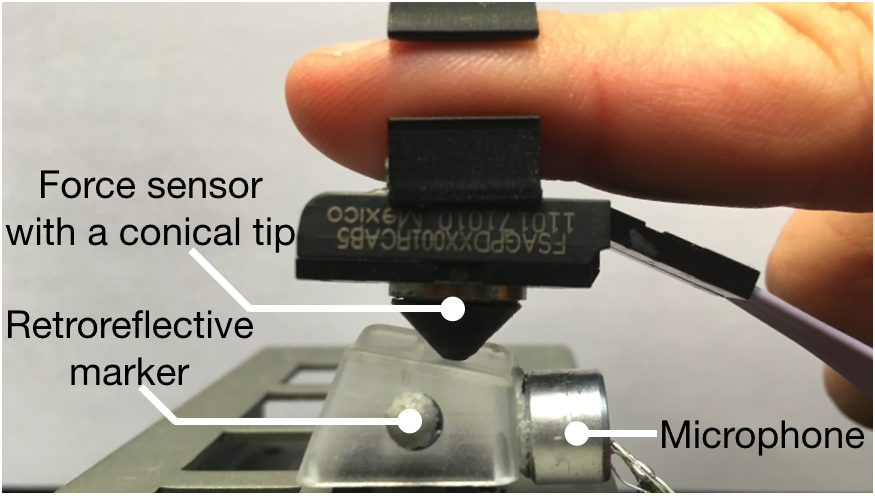}
    \caption{Button capture of an example real button (4~mm tactile button). A force sensor is worn on the fingertip. Reflective markers (for motion tracking) and microphone (for vibration detection) are attached on the keycap.}~\label{fig:capture}
  \end{minipage}
\end{marginfigure}

\begin{marginfigure}[-8pc]
  \begin{minipage}{\marginparwidth}
    \centering
    \includegraphics[width=0.99\marginparwidth]{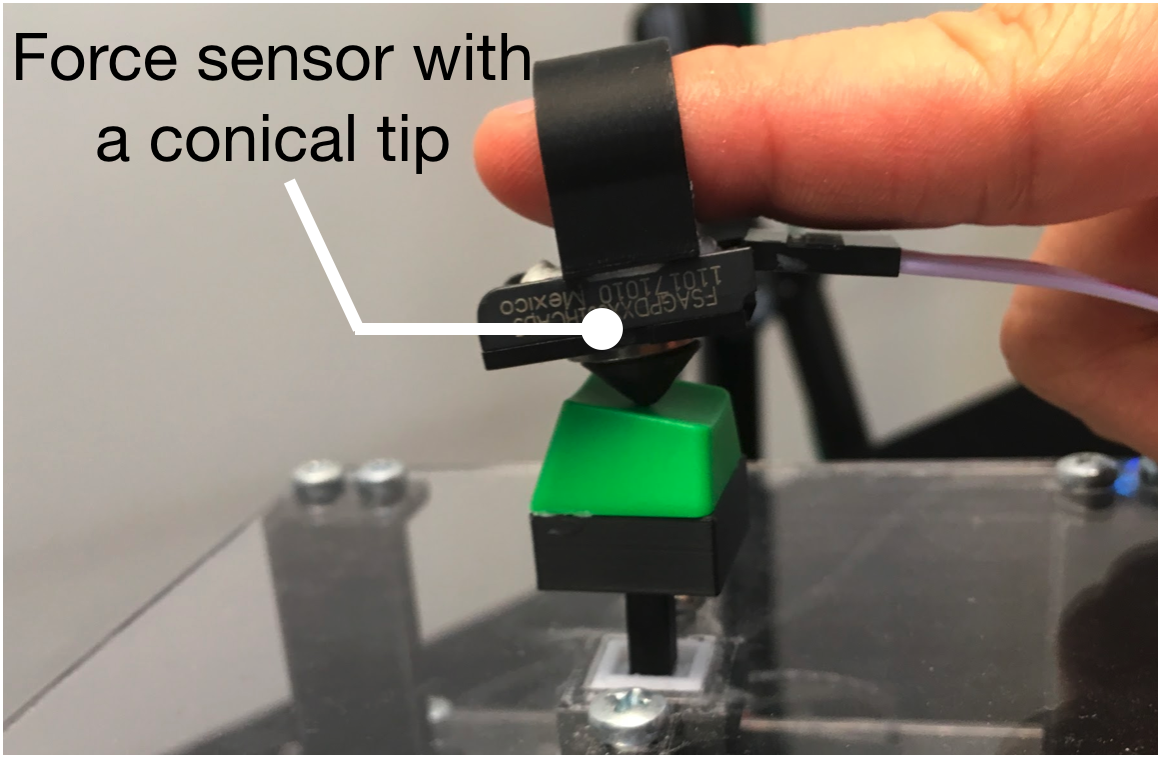}
    \caption{During the iterative compensation, a force sensor is worn on the participant’s fingertip while pressing Press'Em. The sensor gathered force data and sent it to the controller. Which then calculates the errors between reference and sensed values, and tune the actuation signals accordingly until convergence.}~\label{fig:iterative_setup}
  \end{minipage}
\end{marginfigure}

\textbf{Microprocessor design:}
Before simulation, the actuation signals (see Pipeline, step 2) are uploaded to ItsyBitsy and it automatically sets the button travel range. 
During a simulation, the linear sensor constantly sends the reading value to the microprocessor.
A moving-average filter is applied for denoising the reading from the position sensor. 
After the microprocessor has processed the values sent, 
it calculates the current displacement of the button and estimates the user's pressing velocity. 
Then, it determines the corresponding pulse-width modulation (PWM) signal and sends it to the linear force actuator. 
At the displacement where vibration starts, 
the microprocessor sends a command to the Arduino Uno for emitting the vibration.
A high operating frequency is used (1 kHz) for the ItsyBitsy M0 board.

\section{End-to-end Simulation Pipeline}

Press'em replicates tactilities through the following steps: 

\textbf{1. Capture a Button with Different Pressing Velocities:} In previous works, probing machines were used to press buttons with static and slow velocity and summarize the forces as a single-FD model. 
In contrast, in our approach, a user wears a force sensor on their fingertip and presses the button with different velocities (Figure~\ref{fig:capture}). 
Meanwhile, the force, vibration, and displacement data are collected and profiled. 
As plotted in Figure~\ref{fig:fdvvmodel} green lines, the resulting FDVV models show obvious differences between human-pressing curves and the static-machine-press ones.

\begin{marginfigure}[-14pc]
  \begin{minipage}{\marginparwidth}
    \centering
    \includegraphics[width=0.99\marginparwidth]{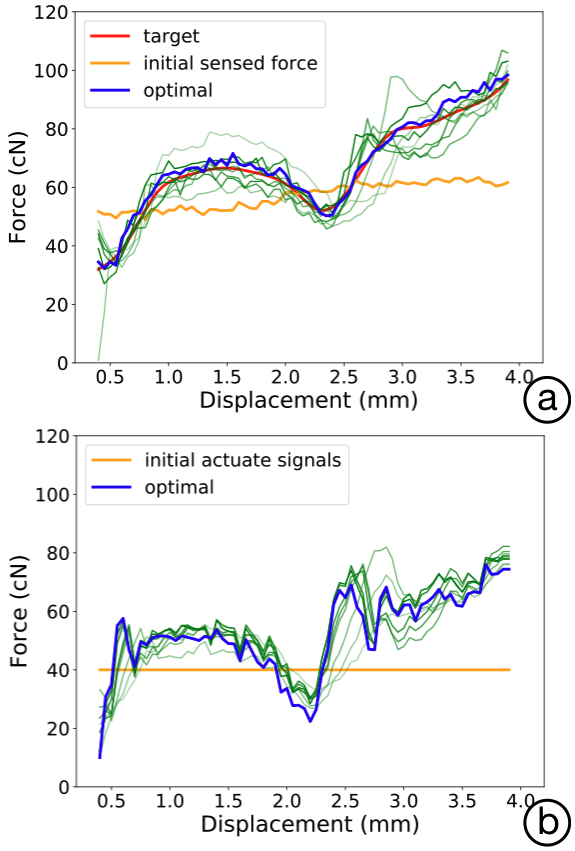}
    \caption{An example iterative compensation process from which we can see: (a) the sensed force from the sensor on the fingertip converges with the reference after compensation is complete, and (b) the actuation signals of the same example starting at a random force level and being gradually tuned.}~\label{fig:signal}
  \end{minipage}
\end{marginfigure}

\textbf{2. Derive the Actuation Signals by Iterative Compensation:} 
Any force actuator has its own transfer function in play that must be canceled out if an FDVV model is to be simulated  correctly.
Thus, we apply \emph{Iterative Compensation} to control the rendered force to keep them aligned with the reference, \textit{i.e.}, the models made in previous step.
The idea is to observe the errors between reference and detected responses at each displacement point of each press, then tune the actuation signals until the error is smaller than a threshold.
Figure~\ref{fig:signal} shows an example with 10 iterations of tuning. Press'Em generates force that's close to the reference with average 2.27 cN error-offset after the iterations. 

\begin{marginfigure}[0pc]
  \begin{minipage}{\marginparwidth}
    \centering
    \includegraphics[width=0.99\marginparwidth]{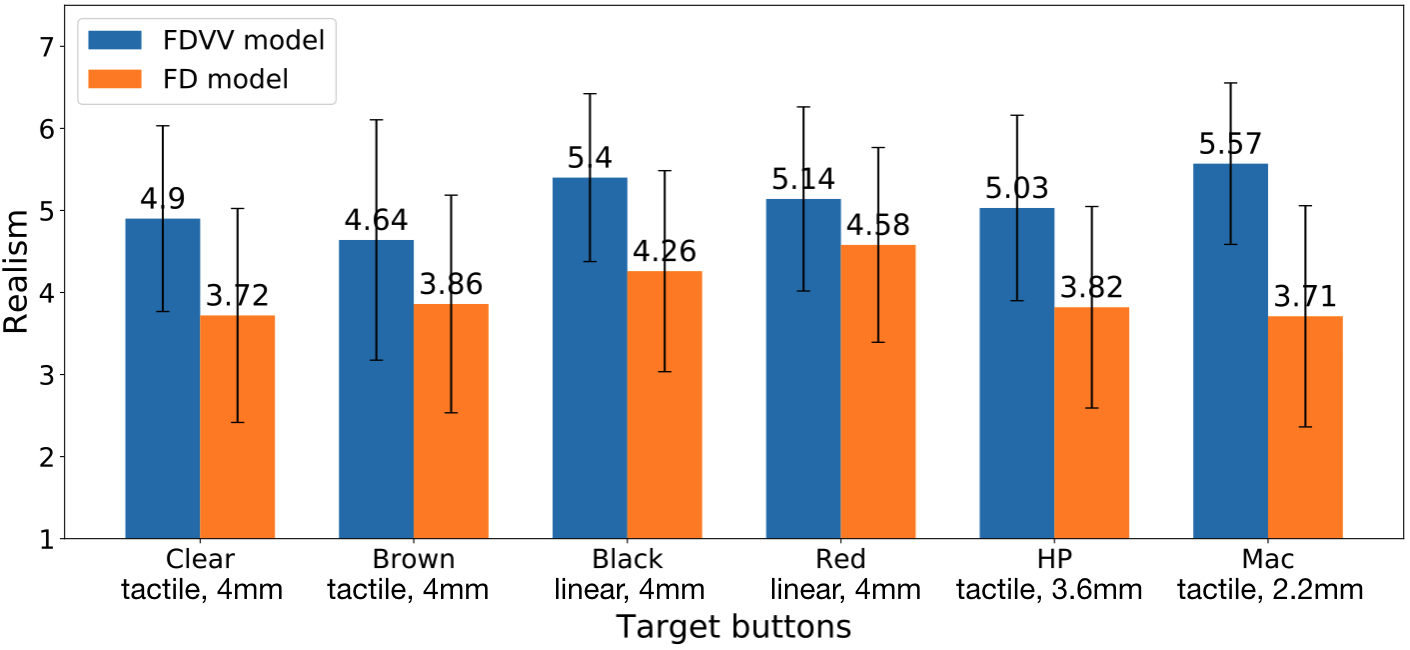}
    \caption{Users in the identity-matching study rated FDVV-based simulations as more realistic than FD simulations. Significant differences were found for all the buttons ($p<0.05$). The target buttons are Cherry MX Clear \& Brown (4mm, tactile), Red \& Black (4mm, linear) switches, HP PR1101U (3.6mm, tactile), and MacBook Pro 2011 (2.2mm, tactile). Error bar is 1 STD.}~\label{fig:results}
  \end{minipage}
\end{marginfigure}

\textbf{3. Real-time Simulate Button Tactilities:} In real-time simulation, everytime the key is pressed, Press'Em detects the displacement and velocity, then send the corresponding force actuation signals and emit recorded vibration that had been derived in the previous steps. 
A 12-participants study was conducted and showed our FDVV models with Press'Em achieves higher perceived realism than traditional FD models to the original buttons (Figure~\ref{fig:results}).

\section{Application: Innovative Buttons}
Press'Em further allows designers to freely explore designs, even those can not be realized by mechanical structures. 

\textit{1. A Fast Tapping Button}: While humans can reach about 4 presses per second in tapping tasks, Press'Em can increase such human capability. The principle is once a press is detected, the button will drop to bottom and return automatically. This could be useful for contents requiring high-frequency and rhythmic tapping, such as music games.

\textit{2. A Dynamically Returning Button}: In certain situations, one button might be desired to avoid fast repetition. Take fighting games as an example, many attacks come with a cooldown time, i.e., the minimum duration before next time using the same skill. Press'Em can render such buttons with dynamic returning time according to the contents.

\textit{3. Rich Vibration Cues}: Press'Em can deliver rich temporal information through continuous vibration cues while being pressed. This interaction can enhance the efficiency of dwell-press applications. For instance, when the shutter button on a camera is pressed and the camera is continuous shooting, the vibration ticks help the user easily count the number of shots via human's haptic channel.

\balance{} 

\bibliographystyle{SIGCHI-Reference-Format}
\bibliography{sample}

\end{document}